# SCAN STATISTIC TAIL PROBABILITY ASSESSMENT BASED ON PROCESS COVARIANCE AND WINDOW SIZE

ANAT REINER-BENAIM

*University of Haifa, Haifa 31905, Israel*

areiner@stat.haifa.ac.il

ABSTRACT

A scan statistic is examined for the purpose of testing the existence of a global peak in a random process with dependent variables of any distribution. The scan statistic tail probability is obtained based on the covariance of the moving sums process, thereby accounting for the spatial nature of the data as well as the size of the searching window. Exact formulas linking this covariance to the window size and the correlation coefficient are developed under general, common and auto covariance structures of the variables in the original process. The implementation and applicability of the formulas are demonstrated on families of multiple processes of t-statistics. A sensitivity analysis provides further insight into the variant interaction of the tail probability with the influence parameters. An R code for the tail probability computation is offered within the supplementary material.

KEYWORDS: scan statistic; covariance structure; global peak detection; moving sums; sequence search.

# 1. INTRODUCTION

A scan statistic is used to identify an unusual cluster, or interval, of events within a random process. If an event is defined as an exceptionally high observation, a cluster of such events may be referred to as a "peak". Around true peaks of a given length, higher positive observations and a larger number of them are expected, compared to false, artifact, peaks (e.g. Hoh and Ott 2000, Keles, Van der Laan, Dudoit and Cawley 2006, Schwartzman, Gavrilov and Adler 2011). Thus, rather than using point-wise testing for which multiple consecutive rejections constitute a single discovery, combining information from neighboring observations to form a single test can increase the power of locating the true peak (Siegmund, Zhang, and Yakir 2011).

The scan statistic records a maximal score among neighborhoods of observations which it scans continuously along a random process. It is formally defined as follows: Let $X_1,...,X_n$ be a sequence of random variables, and let $Y_w(t)=\sum_{i=t}^{t+w-1} X_i$ be a moving sum of *w* consecutive random variables, $1 \le w \le n$, starting from point *t*, $t=1,...,n-w+1$. Then the linear unconditional scan statistic is defined as (Glaz and Balakrishnan 1999):

$$S_w = \max_{1 \le t \le n-w+1} Y_w(t). \tag{1}$$

Many recent applications of the use of this statistic are related to genomic sequences search. For instance, Karlin and Brendel (1992) suggest using (1) on distances between occurrences of a specific marker in a genomic region, in order to detect over-dispersion of the marker, and within score-based analysis of protein and DNA sequences, in which the segment with maximal aggregate score may indicate an anomaly in the sequence composition. They point out that maximizing over sums of measures rather than over single measures increases detection power and tolerance to measurement error. Hoh and Ott (2000) analyze information on marker loci covering a contiguous area of the genome in search of Autism susceptibility genes. They compute single-marker logarithm of odds (LOD) scores based on the genotypes of each marker, and then use the scan statistic to detect effects with a significance level. Chan and Zhang (2007) search for high concentrations of palindromic patterns along the genome, treating the occurrences of the patterns as a marked Poisson process. A scan statistic of binary match records of nonaligned DNA sequences is discussed by Fu, Lou and Chen (in Glaz, Naus and Wallenstein 2001). The motivating example for this paper is related to the detection of peaks within a random process containing point-wise observed measurements. This paper facilitates a scanning test for a continuous random process, by introducing exact formulas to derive the tail probabilities for the statistic. The novelty of the formulas is in linking the tail probability to the covariance matrix of the process, while accounting for the fact that the moving sums of window size $w$ are maximized, rather than the underlying original variables. The formulas can be readily implemented to obtain p-values for multiple processes for any type of statistic using its location and scale parameters. The emerging nature of the relationship between the scan statistic null distribution and the random process distribution is further explored under various window sizes. The next two subsections describe a motivating genetics example and provide background on available approaches.

## 1.1 A Motivating Example: Detecting Peaks in Tiling Array Data

A challenging high-throughput search problem is posed within the work of Juneau, Palm, Miranda et al (2007), who study intronic activity related to meiosis in the yeast genome. Introns are DNA sequences located within genes, and do not code for proteins (as opposed to exons). They may be removed during the mature RNA generation to allow for alternative splicing of the exons, enabling the synthesis of several protein isoforms based on the same gene. While the length of introns and their per-gene occurrences vary between eukaryotic organisms (Mourier and Jeffares 2003), the yeast genome accommodates the activity of only one intron per gene. Juneau et al. (2007) designed an experiment in which the genetic expression attributed to the intronic regions of interest will increase, and thus reveal their locations. Here the interest is in detecting intervals with an increased expression level relative to the baseline level.

Searching for peaks within spatial processes from continuous distributions has been typically implemented for cases where a process could contain many peaks. The strategy taken in these cases is to test each window (Keles et al. 2006 and Huang, Tiwari, Zou et al 2009) or each *local* maximum (Schwartzman et al. 2011) and then account for multiplicity by controlling a proper error criterion. However, if at most one intron can act within a particular yeast gene as in Juneau et al. (2007), this approach is wasteful since it tests many positions within a given gene while at most one of them can be a peak. Instead, a single per-gene test for the existence of a peak within the gene can be employed for each gene separately, using a scan statistic as defined in (1), with a window size proportional to the length of the peak searched for. In this case, the null hypothesis states that all observations come from identically distributed random variables with a mean parameter $\theta_0$ attributed to their null distribution, denoted here by $F_X(x;\theta_0)$. Assuming a single peak in the process, hence referred to as a *global* peak, the alternative hypothesis states that for a given window size *w*, there exists some $1 \le k \le n-w+1$ for which the consecutive variables $X_k,...,X_{k+w-1}$ are distributed with $\theta_k,...,\theta_{k+w-1} > \theta_0$. The moving sums statistic $Y_w(t)$ reflects the combined elevation level of the neighboring observed expression levels, and its maximum $S_w$ captures the location of the potential peak. Assuming there are $m$ genes within which a peak is searched for, the analysis will yield a series of scan statistics $S_{w,1},...,S_{w,\mathrm{m}}$ for which the corresponding tail probabilities under the null should be evaluated and adjusted for multiplicity.

## 1.2 Scan Statistic Tail Probability under Dependence

Let $F_S^0(s;w)$ be the distribution of $S_w$ corresponding to the null hypothesis. The tail probability $\Pr_{H_0}(S_w > s) \equiv \Pr_{F_S^0}(S_w > s)$ must be evaluated in order to test the above null hypothesis. Approximations for this probability were offered for Gaussian processes by Siegmund (1988) and Kim and Siegmund (1989), later extended by Loader (1991) to Poisson processes, and by Woodroofe (1976). Other results, mostly for the case of i.i.d. Bernoulli trials, are documented in Glaz and Balakrishnan (1999) and in Glaz, Naus and Wallenstein (2001). In particular, exact formulas have been obtained for i.i.d. Bernoulli processes (Naus 1974) and general 0-1 trials (Kourtas and Alexandrou 1995). Poisson-type approximations for i.i.d. variables have been developed by Darling and Waterman (1986) and Goldstein and Waterman (1992), while product-type approximations have been developed by Naus (1982) and Glaz and Naus (1991). These approximations are refined in Glaz and Balakrishnan (1999) for the case of i.i.d variables. Product-type inequalities have been introduced by Glaz and Naus (1991) and Bonferroni type inequalities have been considered by Glaz and Naus (1991) and Chen (1998). For the case of dependent variables, Dembo and Karlin (1992) offered limit distributions for Markov-dependent alphabetical sequences.

Since moving sums may be thought of as a smoothed version of the observations (with the averaging division by $w$ neglected), the scan statistic can be regarded as a maximum over a continuous process. Extreme value theory provides exact formulas for the tail probability of the maximum of a continuous process with any covariance matrix (Leadbetter, Lindgren and Rootzen 1983), based on the expected Euler Characteristics of the smoothed statistics for a given threshold (Rice 1945). Random field theory extends the concept for a smoothed statistical "map" containing multidimensional statistical endpoints (Adler and Taylor 2007, and Taylor and Worsley 2007). The covariance function of the process and its differentiability properties play a critical role in the resulting formulation. However, this approach does not account for the underlying observations of each sum, and is thus equivalent to the case of window size $w=1$, namely taking a maximum over the original process. For a general $w$, such that the original process is transformed into moving sums on which the maximum is taken, this paper introduces a computational scheme that uses the information provided by the covariance function of the original process along with the window size in order to obtain the exact scan statistic tail probabilities. The core of the computational work is finding the variance and covariance functions of the moving sums in the process, which once known, serve in evaluating the requested probabilities. The formulas can be applied to any distribution of the variables in the process, and are effective in obtaining the p-values for multiple processes, each corresponding to a single hypothesis.

In Section 2 of this paper, exact and readily implemented formulas for the variance and covariance are introduced, first for a general covariance structure of the original process and then for the cases of common correlation and auto-correlation structures. While it is assumed that the exact covariance matrix of the process is known, in practice it must be estimated, and this paper does not deal with the resultant inferential properties of the scanning test. The derivation of the formulas is presented in Appendix A. Implementation for multiple processes of t-statistics is demonstrated in Section 3. In Section 4, the obtained formulas are further explored through studying the sensitivity of the resulting probabilities to the window size and correlation coefficient. A verification analysis of the analytical results, presented in Appendix B, compares the obtained probabilities to Monte Carlo simulation estimates. An R function employing the obtained formulas as well as a demonstration of using the formulas on a single process are provided as supplementary for this paper. The paper concludes with a discussion of the results, and their effective incorporation into the analysis. Further applications as well as practical problems are stretched to motivate further study on the implementation of the proposed computational results.

## 2. THE COVARIANCE FUNCTION OF THE MOVING SUMS

For an observed scan statistic *s*, clearly

$$\begin{aligned}\Pr_{H_0}\left(S_w \le s\right) &= \Pr_{H_0}\left(\max\left[Y_w(1),\ldots,Y_w(n-w+1)\right] \le s\right) \\ &= \Pr_{H_0}\left[Y_w(1) \le s,\ldots,Y_w(n-w+1) \le s\right],\end{aligned}$$

which is the null joint distribution of the moving sums at the $(n-w+1)$–dimensional coordinate $s,\ldots,s$, obtained by integrating the corresponding joint density function,

$$\int_{-\infty}^{s} \dots \int_{-\infty}^{s} f_{Y(1),\dots,Y(n-w+1)}(s,\dots,s;\theta_0,w,n)\; dY(1)\dots dY(n-w+1). \tag{2}$$

Once the density is known, the integral in (2) can be approximated, and the requested tail probability is its complement. Numerical approximation for the integral are offered by Genz 1992, 1993 for the Gaussian case and by Genz and Bretz 2009 for the t distribution, and can be implemented using the mvtnorm R package (Genz et al. 2011). But first, the density must be found. While the assessment of the mean vector for the moving sums is typically trivial, as in the normal and t cases, obtaining the covariance matrix of the sums requires computational effort.

Consider the random process $X_1,\dots,X_n$, where each $X_i$ is distributed with a mean parameter $\theta_0$ under the null hypothesis. Let $\mathrm{var}(X_i)=\sigma^2$ for all $i$ and let $cor(X_i,X_j)=\rho_{ij}$ for all $i \neq j$. For $t=1,\dots,n+w-1$, $Y_w(t)$ has variance $\sigma_{Y(t)}$, and denote by $\Sigma_Y$ the covariance matrix of the moving sums process $Y_w(1),\dots,Y_w(n+w-1)$. The following theorem links the covariance between any two moving sums with the covariance of the variables in the process.

*Theorem 1.* Let $X_1,\dots,X_n$ be a stationary process of homoscedastic random variables with covariance matrix $\Sigma_X$, where $\mathrm{var}(X_i)=\sigma^2$ and $cor(X_i,X_j)=\rho_{ij}$ for all $1 \le i \neq j \le n$. Let $Y_w(t)=\sum_{i=t}^{t+w-1} X_i$ and let *g* be a positive integer, $1 \le g \le n-1$. Then the covariance between $Y(t)$ and $Y(t+g)$ is given by

$$\mathrm{cov}\left[Y(t),Y(t+g)\right]=\sigma^2\left[\sum_{i<j\subset A}\rho_{ij}+\sum_{i<j\subset C}\rho_{ij}+\sum_{\substack{i\subset A\\ j\subset C}}\rho_{ij}\right.$$
$$\left.+2\left(2\sum_{i<j\subset B}\rho_{ij}+\sum_{\substack{i\subset A\\ j\subset B}}\rho_{ij}+\sum_{\substack{i\subset B\\ j\subset C}}\rho_{ij}\right)-\sum_{i<j\subset[t,t+w-1]}\rho_{ij}-\sum_{i<j\subset[t+g,t+g+w-1]}\rho_{ij}+w-g\right] \tag{3}$$

where $A=[t,t+g-1]$, $B=[t+g,t+w-1]$ and $C=[t+w,t+g+w-1]$. A proof is given in Appendix A.1, which in addition shows that in the case of non-overlapping windows such that $g \ge w$, (3) simplifies to

$$\mathrm{cov}\left[Y(t),Y(t+g)\right]=\sigma^2\sum_{\substack{i\subset[t,t+w-1]\\ j\subset[t+g,t+g+w-1]}}\rho_{ij}. \tag{4}$$

The next two corollaries apply to special cases of $\Sigma_X$.

*Corollary 1.* For the process in Theorem 1, assume auto-correlation between $X_1,...,X_n$, such that $\rho_{ij} = \rho^{|j-i|}$ for any $i \neq j$. Then

$$\operatorname{cov}\left[Y(t),Y(t+g)\right]=\sigma^2\left[\sum_{i=1}^{g}\sum_{j=1}^{g}\rho^{|w+j-i|}+2\left(2\sum_{i=1}^{w-g-1}(w-g-i)\rho^i+\sum_{i=1}^{g}\sum_{j=1}^{w-g}\rho^{|g+j-i|}\right.\right.$$
$$\left.\left.+\sum_{i=1}^{w-g}\sum_{j=1}^{g}\rho^{|w-g+j-i|}+\sum_{i=1}^{g-1}(g-i)\rho^i-\sum_{i=1}^{w-1}(w-i)\rho^i\right)+w-g\right] \tag{5}$$

A proof is given in Appendix A.2, which in addition shows that in the case of non-overlapping windows such that $g \geq w$, (5) simplifies to

$$\operatorname{cov}\left[Y(t),Y(t+g)\right]=\sigma^2\sum_{i=1}^{w}\sum_{j=1}^{w}\rho^{|g+j-i|} \tag{6}$$

*Corollary 2.* For the process in Theorem 1, assume a common correlation between $X_1,...,X_n$, such that $\rho_{ij} = \rho$, $\frac{-1}{n-1} \leq \rho \leq 1$, for any $i \neq j$. Then

$$\operatorname{cov}\left[Y(t),Y(t+g)\right]=\sigma^2\left[\rho\left\{g(2g-1)+2(w-g)(w+g-1)-w(w-1)\right\}+w-g\right]. \tag{7}$$

A proof is given in Appendix A.3, which in addition shows that in the case of non-overlapping windows such that $g \geq w$, (7) simplifies to

$$\operatorname{cov}\left[Y(t),Y(t+g)\right]=\sigma^2\sum_{\substack{i\subset[t,t+w-1]\\ j\subset[t+g,t+g+w-1]}}\rho_{ij}=\sigma^2 w^2\rho. \tag{8}$$

The reader is further referred to Appendix B, which includes a verification study for the formulas given in this section using tail-probability estimates based on simulation. The results presented there confirm the equivalence of the exact and estimated covariance matrices for an arbitrary covariance structure and for the common and auto-covariance structures.

## 3. IMPLEMENTING SCAN TESTING FOR THE T-DISTRIBUTION

This section implements the proposed formulas for multiple processes containing t-distributed variables. t-statistics are often used in practice for several types of popular analyses, such as comparing the means of two populations or fitting a linear regression model. In some cases, it may be relevant to perform point-wise t-testing across time or location, and then to scan them for regions of high level. For instance, in genomic research, it is sometimes of interest to identify regional elevations relative to a null process. For the intron searching problem studied by Juneau et al. (2007) and referred to in the Introduction, the expression levels, typically log-transformed into an approximate normal scale, but replicated a small number of times, may be compared point-wise between the strains using t-statistics. The technology used to measure the expression levels is the tiling microarray, which evaluates expression of overlapping or adjacent short sequences along the genome. Due to this overlap or adjacency, as well as the geographical nearness of the probes on the array, the measurements, as well as the calculated statistics, have a spatial structure and may thus be auto-correlated (e.g. Reiner, Yekutieli and Benjamini 2003). The t-statistics may be calculated for the whole genome, and then the long series of statistics is divided into segments by genes, such that each gene can be separately scanned for the existence of an intron, using a scan statistic. Then, the p-value may be calculated for each gene based on (3), (5) or (7), and all resulting p-values are corrected for multiplicity.

Even though the t-statistics may be transformed into normal scores (see Efron, 2007, 2010), the scan statistics' p-values may not be readily obtained based on these scores using the previous approximations for the Gaussian normal case (Woodroofe, 1976, Siegmund, 1988 and Kim and Siegmund, 1989). The reason is that the covariance structure of the normal scores is different from the covariance structure of the original t-statistics, as demonstrated in Table 1 for simulated data. Unless the variance of the t-statistics is 1, the variance of the normal scores is different. In addition, unless the correlation coefficient between the t-statistics is 0, the correlation coefficient between the normal scores is slightly less than the original coefficient. Thus, if the original t-statistic covariance parameters are plugged in, formulas (3), (5) and (7) will generate an erroneous covariance matrix for the moving sums of the normal scores. Yet, these formulas work well when using the t-statistics, as can be seen in the resulting p-values presented in Table 2 for several covariance settings.

In order to demonstrate the applicability of the formulas for a searching problem within a genomic data, statistics were generated according to a data structure typical for tiling array output. Auto-correlated t-statistic series of lengths distributed similarly to the gene lengths in the yeast genome (accessed through the SGD database site http://www.yeastgenome.org/) were sampled, using varying configurations of correlation coefficient, effect length and effect height. The number of simulated series was 6000, which is around the actual number of genes in the yeast genome. 4% (240) of the genes had an "effect" - a region with elevated t-statistics. The p-values for all 6000 observed gene-wise scan statistics were calculated using the covariance matrix obtained by (5) and evaluating the integral in (2). As can be seen in Figure 1 for a parameter configuration typical for tiling array data, the observed p-values conform with a $U[0,1]$ distribution, as expected for the case of no effect, and become smaller than expected under $U[0,1]$ for genes containing an effect. This behavior was consistent for all other configurations examined.

Table 1: Covariance parameters for the t-statistics and normal statistics.
The t-statistics were samples from a multivariate t-distribution with 7 degrees of freedom and preset variance $\sigma_t$ and correlation coefficient $\rho_t$, while the corresponding parameters for the normal scores, $\sigma_s$ and $\rho_s$, were estimated from the transformed statistics.
The process length was 100. 100 simulations of 1000 processes were used.

| Covariance structure | $\sigma_t$ | $\rho_t$ | $\hat{\sigma}_s$ | $\hat{\rho}_s$ |
|---|---|---|---|---|
| Common | 1 | 0 | 1.002 (0.001) | 0 |
| | | 0.25 | 0.999 (0.002) | 0.248 (0.001) |
| | | 0.5 | 1.002 (0.002) | 0.497 (0.001) |
| | | 0.75 | 0.999 (0.003) | 0.747 (0.001) |
| | 2 | 0 | 2.815 (0.003) | 0 |
| | | 0.25 | 2.808 (0.004) | 0.242 (0.001) |
| | | 0.5 | 2.815 (0.006) | 0.489 (0.001) |
| | | 0.75 | 2.826 (0.008) | 0.742 (0.001) |
| | 4 | 0 | 6.388 (0.008) | 0 |
| | | 0.25 | 6.393 (0.007) | 0.235 (0.0009) |
| | | 0.5 | 6.39 (0.009) | 0.476 (0.001) |
| | | 0.75 | 6.407 (0.013) | 0.728 (0.0009) |
| Auto | 1 | 0.25 | 0.997 (0.0018) | 0.248 (0.0003) |
| | | 0.5 | 1.002 (0.0017) | 0.497 (0.0003) |
| | | 0.75 | 0.998 (0.0018) | 0.747 (0.0002) |
| | 2 | 0.25 | 2.814 (0.0038) | 0.243 (0.0003) |
| | | 0.5 | 2.815 (0.004) | 0.49 (0.0003) |
| | | 0.75 | 2.82 (0.004) | 0.74 (0.0002) |
| | 4 | 0.25 | 6.381 (0.006) | 0.243 (0.0003) |
| | | 0.5 | 6.388 (0.007) | 0.476 (0.0002) |
| | | 0.75 | 6.384 (0.007) | 0.726 (0.0002) |

Table 2: Exact and estimated tail probability for scan statistics of t processes. Observations were sampled from a multivariate t-distribution with 7 degrees of freedom and scale parameter $\sigma_t = 4$. The process length was 7 and window size was 3 . The simulation was repeated with $N$=1000 and $J$=500. In parenthesis - standard error for the estimate. Equality of exact and estimated results was consistent for other values of process parameters. The tail probability is calculated for an "observed" scan statistic $s$=3

| Covariance structure | $\rho$ | $p_w(s)$ | $\hat{p}_w(s)$ |
|---|---|---|---|
| Common | 0 | 0.71246 | 0.7132 (0.0006) |
| | 0.25 | 0.6202 | 0.6216 (0.0007) |
| | 0.5 | 0.5600 | 0.5608 (0.0008) |
| | 0.75 | 0.5056 | 0.5060 (0.0007) |
| Auto | 0.25 | 0.6998 | 0.6997 (0.0007) |
| | 0.5 | 0.6693 | 0.6689 (0.0007) |
| | 0.75 | 0.6080 | 0.6093 (0.0007) |

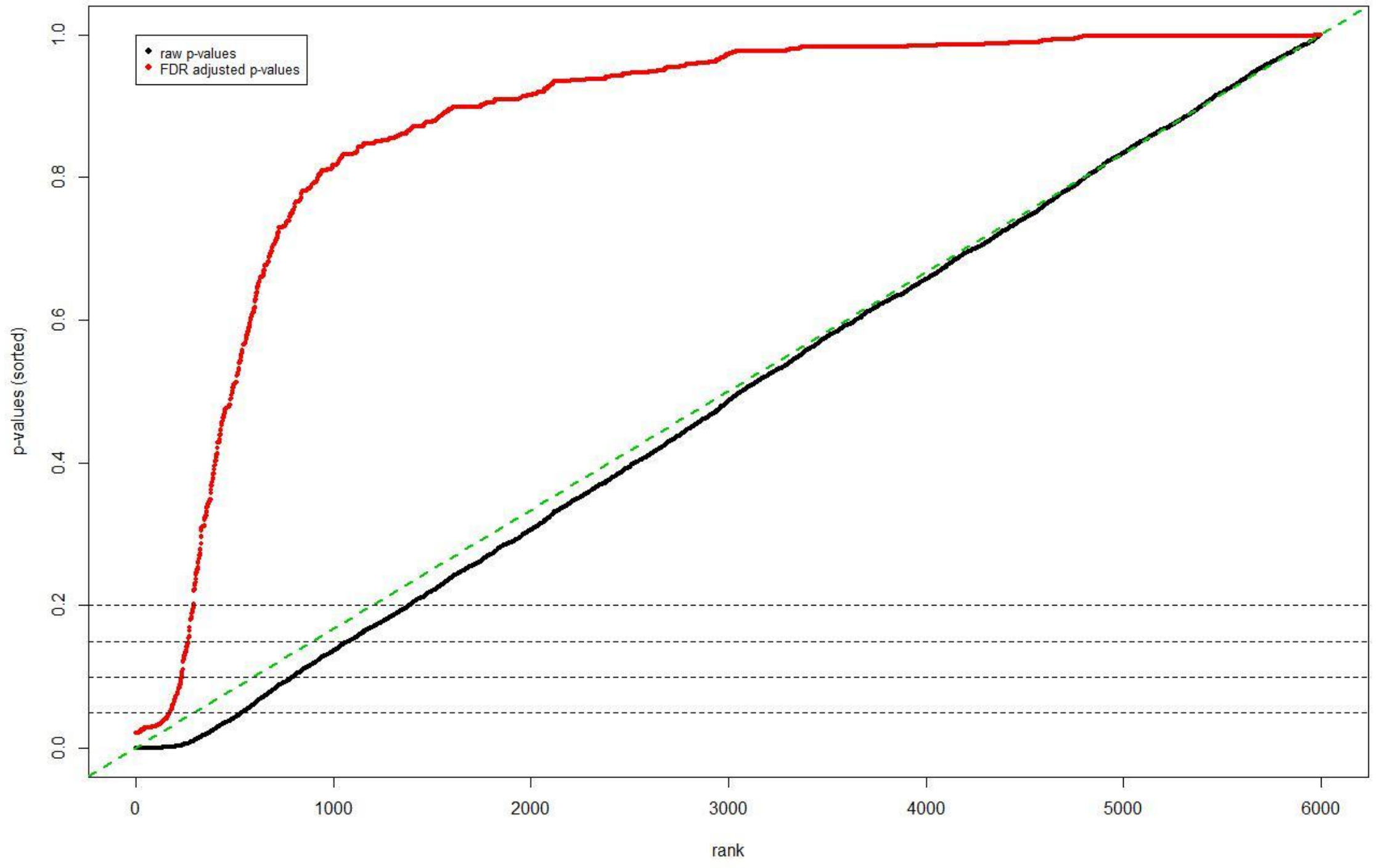

Figure 1. P-values (black) and FDR-adjusted p-values (red) for 6000 scan statistics based simulated auto-correlative t-processes with 7 degrees of freedom, of which 240 (4%) had an effect. Dotted green line marks $U[0,1]$. Process lengths were sampled from $\{100, 500, 1000\}$ with probabilities $\{0.4, 0.5, 0.1\}$, respectively, and effect height was sampled from $\{12,14,16,18\}$ with equal probabilities. $\rho = 0.2$, $\sigma = 4$, effect length is 10 and window size is 10. The false discovery proportion was 0.034 and power was 0.7.

## 4. SENSITIVITY ANALYSIS

This section provides a further insight into the obtained relations between the covariance of the sums and the process and scan statistics parameters, first by evaluating the covariance using the formulas introduced in Section 2. It also examines the resulting scan statistics distribution and the corresponding tail probabilities by approximating the integral in (2) for the normal case.

Formulas (5) and (7) were used for auto-covariance structure and for common covariance structure, respectively, to compute the covariance for the case of overlapping windows. Overlap is represented by the gap (distance) between the starting points of the two windows, given as a proportion of the total window size. Thus the larger the proportion, the small that gap, and when $g = w$, the windows are adjacent.

Figure 2 visualizes the contributions of the correlation coefficient and overlap size to the total covariance, for two different window sizes. In the case of common correlation (Figure 2a), while the covariance increases linearly in $\rho$, an overlap will contribute a further increase. Furthermore, clearly at $\rho = 0$, only overlapping windows will have covariance, and as $\rho$ increases, the relative contribution of the overlap diminishes until a complete vanish at $\rho = 1$. In the case of auto-covariance (Figure 2b), the covariance accelerates in $\rho$. The effect of overlap size increases in $\rho$ until some point from which it starts to diminish.

Figure 3 provides another perspective on the contribution of the window size to the covariance of the sums. Generally, the effect of window size increases in $\rho$. In the case of common covariance structure (Figure 3a), the effect of window size is only vaguely interacted with the overlap size. However, in the case of auto-correlation (Figure 3b), the effect of window size increases in overlap size. For adjacent windows, the window size takes an effect only starting from a certain value of $\rho$.

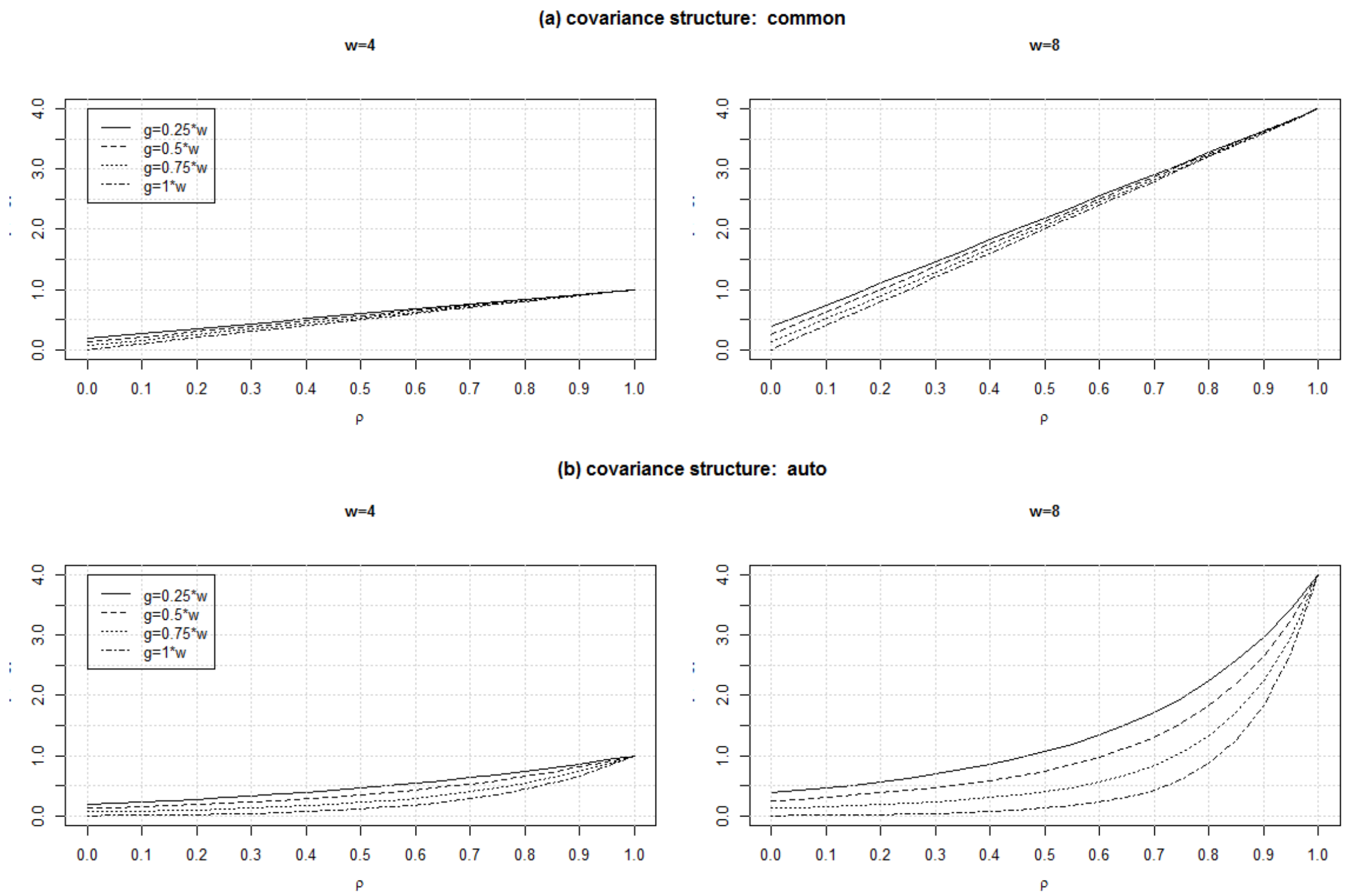

Figure 2. Covariance of window sums of normal random variables vs. the correlation coefficient, by window size. (a) Common process covariance structure (b) Auto process covariance structure. Process length is 16. The correlation coefficient studied is positive, since for common correlation only negative values that are very close to zero are possible under positive definiteness of the process covariance matrix.

Next, the effect of the process and scan statistic parameters on the scan statistic null distribution and its corresponding tail probabilities was examined. A preliminary impression as to the impact of these parameters can be gathered by first observing their effect on the steadiness and periodicity of the underlying normal process, as demonstrated in Figure 4. A Process with a common covariance structure (Figure 4a) tends to be rigid around the level of its first variable, since the correlation between any pair of variables is the same regardless of the distance between them. As the correlation coefficient increases, this pattern becomes more prominent. By contrast, a process with auto-covariance structure (Figure 4b) is easier to deviate from a value observed at a given position, since the correlation between any two variables declines as the distance between them increases. Thus for a given correlation coefficient, the cycles formed under an auto-covariance structure have higher amplitudes, compared to the common case. A higher correlation coefficient leads to longer wave lengths, since the correlation with a given variable is initially high and thus takes more distance to vanish.

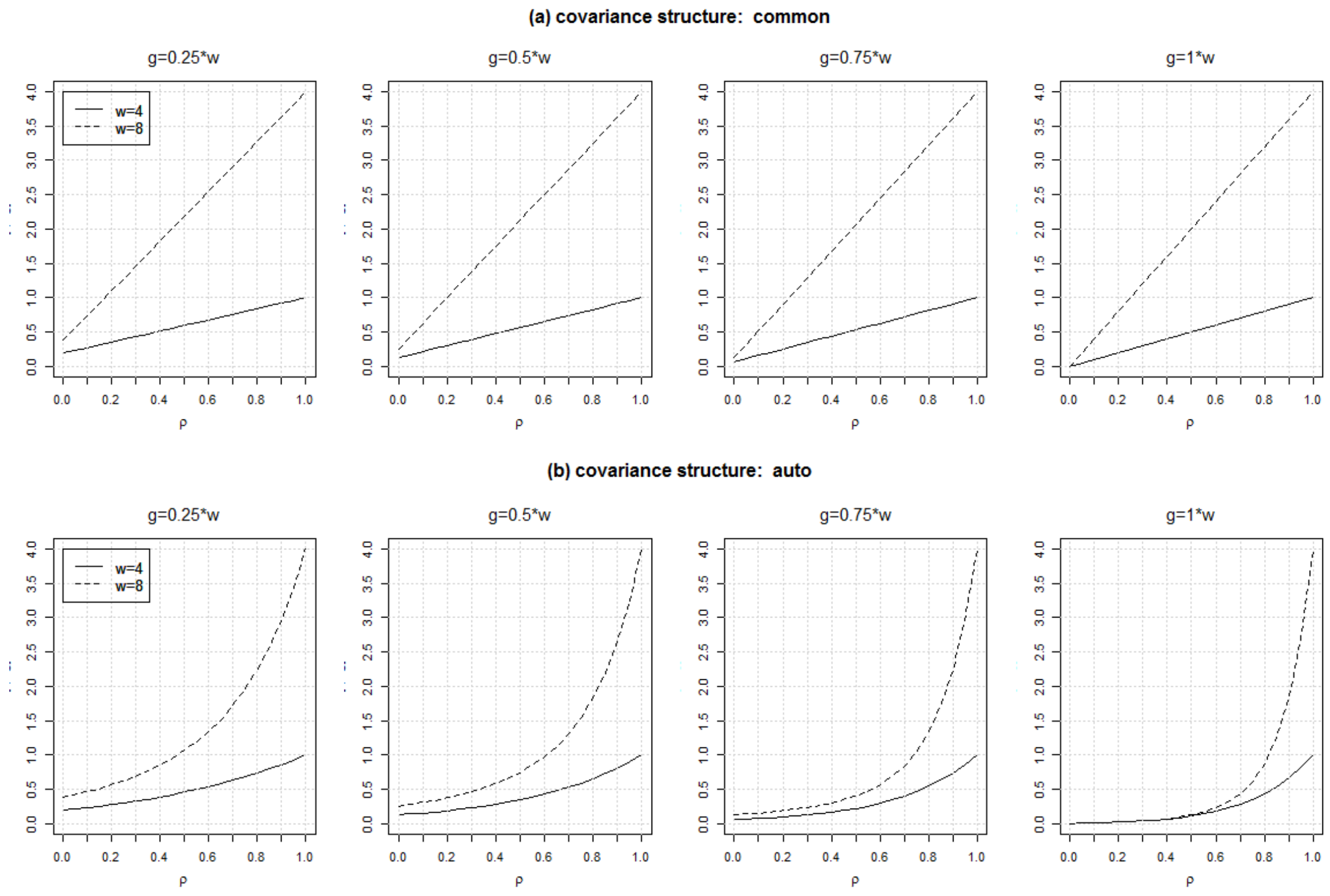

Figure 3. Covariance of window sums of normal random variables vs. the correlation coefficient, by overlap size. (a) Common process covariance structure (b) Auto process covariance structure. Process length is 16. The correlation coefficient studied is positive, since for common correlation only negative values that are very close to zero are possible under positive definiteness of the process covariance matrix.

Simulated standard normal processes that do not contain "peaks" served to empirically estimate the scan statistic distribution characteristics for different configurations of covariance structure, correlation coefficient and window size. In addition, based on the exact moving sums covariance function, scan statistic tail probabilities were approximated for each configuration. The scan statistic mean, standard deviation, skewness, kurtosis and tail probability for a given observed value were calculated for a normal process of length 16 and plotted in Figure 5 and Figure 6 as functions of the influence parameters, under common covariance and auto-covariance structures, respectively.

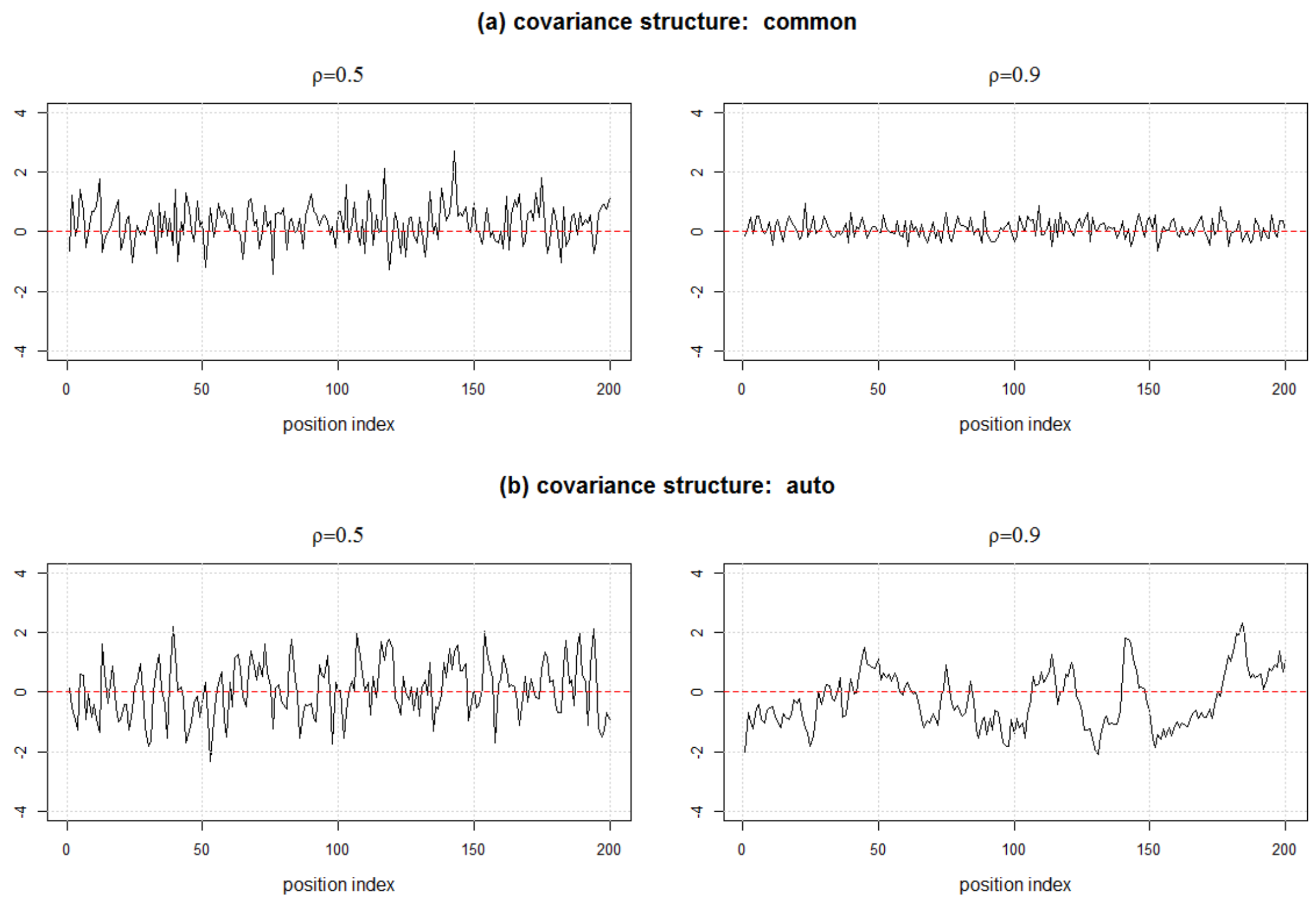

Figure 4. Simulated standard normal processes. Process length is 200.

The mean in the case of common covariance decreases in $\rho$, due to the parallel decline in the potential of obtaining high observations in the underlying process. In the case of auto-covariance, the mean increases in $\rho$ up to some $\rho$, and then start decreasing. As $\rho$ increases, the standard deviation for both covariance structures increases, decelerating in $\rho$ in the case of common covariance, and accelerating in the case of auto-covariance. As $\rho$ becomes smaller, the distribution becomes somewhat more right skewed and its peak becomes somewhat sharper relative to its tails. The tail probability for an observed scan statistic decreases in $\rho$ in the case of common correlation, and has a maximum at some $\rho$ in the case of auto-correlation.

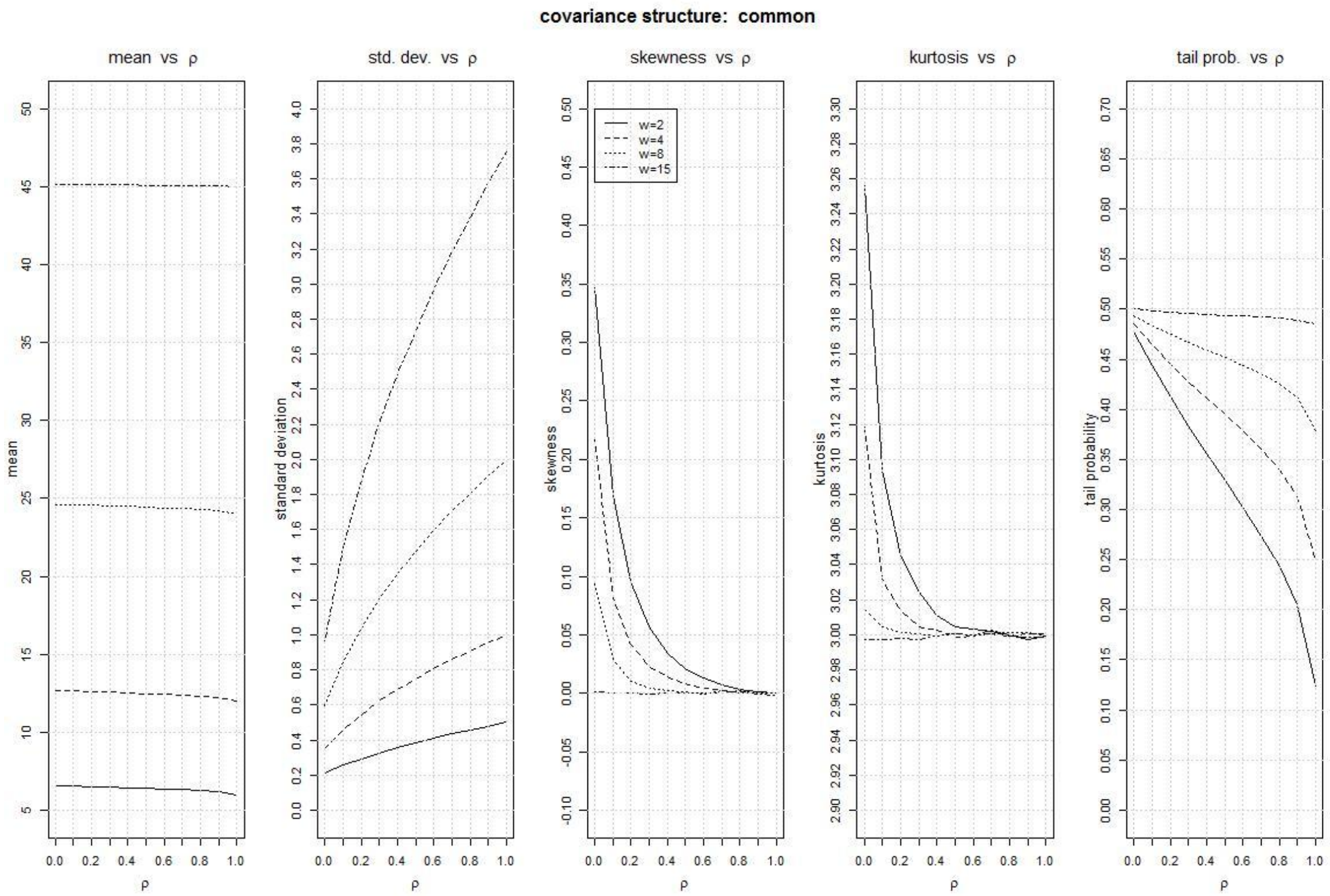

Figure 5: Scan statistic distribution simulation estimates and approximated tail probability vs. the correlation coefficient, by window size - common covariance structure. Process length is 16 and process distribution parameters are $\mu = 3$ $\sigma = 0.25$. The correlation coefficient studied here is between zero and one, since it could obtain only nearly zero negative values under positive definiteness of the process covariance matrix. For a given window size, the tail probability is calculated for an observed scan statistic equal to the mean under no correlation.

The smaller the window size, the further the distribution of the resulting scan statistic from normality, as can be seen by the increased deviation of the skewness and kurtosis away from the values attributed to the normal distribution. Clearly, as the window size approaches the process length $n$ (shown for $w = n - 1 = 15$), the scan statistic approaches $Y_w(1) = \sum_{i=1}^{w} X_i$, the sum of the normal variables in the process, and its distribution approaches normality due to the central limit theorem.

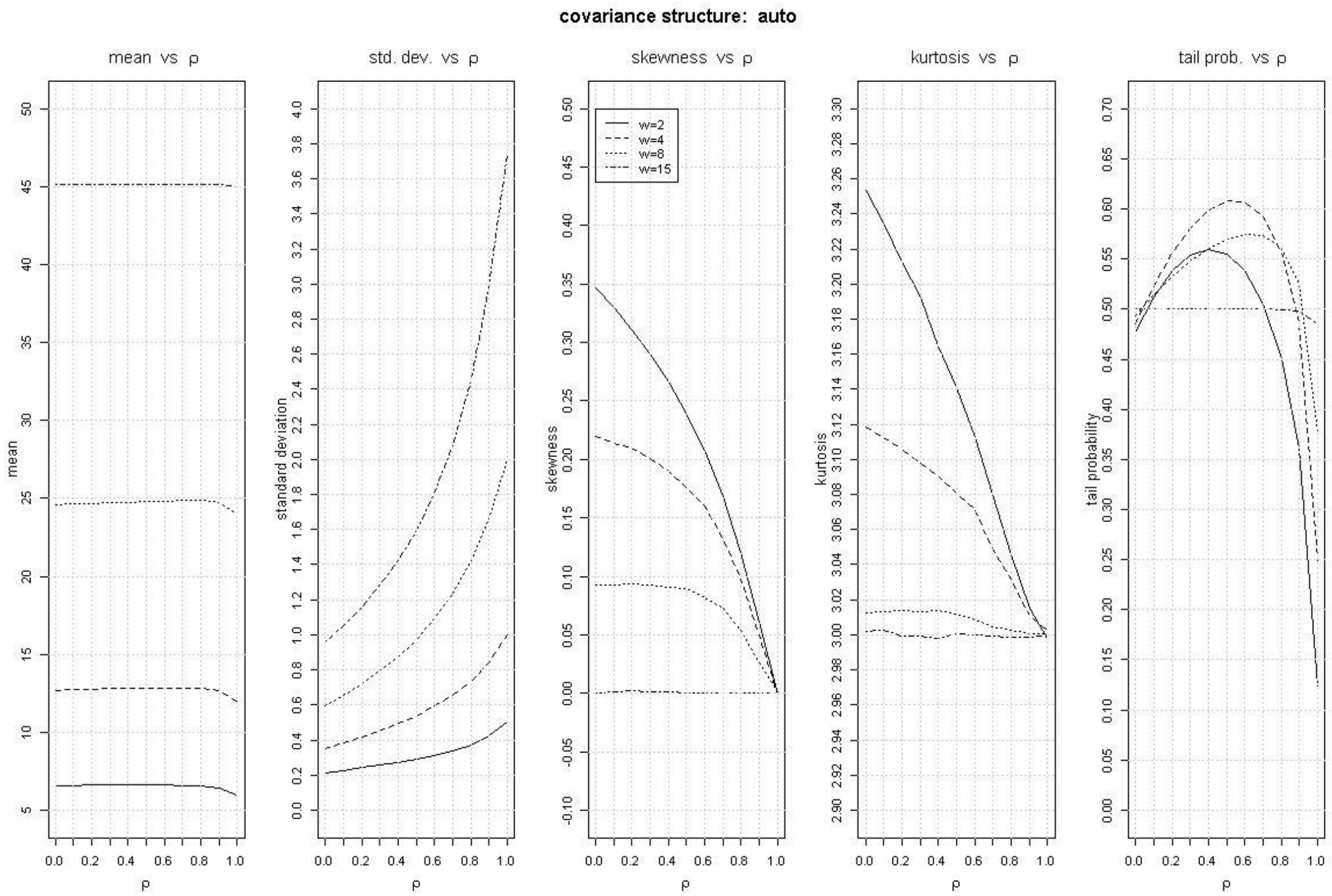

Figure 6: Scan statistic distribution simulation estimates and approximated tail probability vs. the correlation coefficient, by window size – auto-covariance structure. Process length is 16 and process distribution parameters are $\mu = 3 \quad \sigma = 0.25$. The correlation coefficient studied here is between zero and one, since it could obtain only nearly zero negative values under positive definiteness of the process covariance matrix. For a given window size, the tail probability is calculated for an observed scan statistic equal to the mean under no correlation.

## 5. DISCUSSION

This paper derives exact formulas that link the covariance of a random process of any distribution to the distribution of a scan statistic from that process. Based on the covariance structure and the window size w, the covariance matrix of a moving sum can be calculated and then an approximation algorithm may be used to compute the tail probability of the corresponding scan statistic. Simulated scan statistics illustrated the sensitivity of the scan statistic distribution to the correlation coefficient in location, dispersion, symmetry and peakedness. The corresponding tail probabilities obtained using the proposed moving sum covariance formulas are considerably affected by the correlation coefficient of the underlying process and the window size, and the form of the effect along with the monotonicity in $\rho$ are subjected to the covariance structure.

The proposed computational approach is suitable when a certain structure can be assumed and the correlation coefficient can be estimated, sparing the necessity of simulating processes in order to estimate the tail probability for an observed scan statistic (as demonstrated in Appendix B). Obtaining a good estimate of $\rho$ is not trivial, and the stationarity characteristics of the genome-wide sequence should be accounted for. If the sequence is treated as infinite due to its immense length, stationarity is possible, in which case the gene-wise sequences can be treated as its subsets in order to separately estimate their gene-wise correlation coefficients. Yet if the genes can be assumed to emerge from the same covariance structure, as may be the case for the tiling array expression data, their data can be combined in order to provide a pooled estimate for the correlation coefficient. As some evidence for non-stationarity of genomic sequences has accumulated (Bouaynaya and Schonfeld 2008 and Adak 1998), models for the non-startionary case have been proposed as well. See Zelinski, Bouaynaya, Schonfeld and O'Neill (2008) for a discussion on stationarity of genomic expression sequences, a proposed method for the non-stationary case and a review of other available methods. For the obtained vector of process-wise marginal tail probabilities, a multiple testing procedure can be implemented for the control of the chosen type I error criterion (e.g. Reiner et al. 2003).

While the emphasis in this paper is on the scan statistic tail probability computation, the implementation of the scan statistic approach may be further investigated by examining various peak detection problems, particularly for those related to genomic information. Such a study may be helpful in exploring the advantages of using a scan statistic in terms of error rate control and detection power, especially in light of other proposed techniques for analyzing such data. One optional technique may be segmentation, which relies on abrupt changes in the process level and typically assumes piecewise continuous signals. It has been used for estimating DNA copy number, as reviewed in Chen, Xing and Zhang (2011), who also extend this approach to account for fractional copy changes using a continuous-state hidden Markov model. Huber, Toedling and Steinmetz (2006) proposed a segmentation algorithm for detecting transcript boundaries which they implement on yeast expression data produced by tiling arrays (David et al. 2006). This method is shown to be more effective in accurately estimating the boundary locations compared to a sliding window smoother. However, for both applications the interest is in identifying all change points along the genomic sequence, while the scan statistic discussed in this paper is aimed to test the existence of a single peak within a process. Nevertheless, if a peak is identified by a scan statistic, a segmentation model may be fitted post-hoc in order to identify its boundaries. Thus, the boundary estimation may be regarded as hypotheses which are tested conditionally on the rejection of the peak hypotheses. If there are multiple processes, the peak hypotheses and boundary hypotheses form a hierarchical testing flow and then the multiplicity of tests may be confronted by implementing an FDR controlling procedure which is suitable for such an organization of tests (Yekutieli et al. 2006, Reiner-Benaim et al. 2007).

## APPENDIX A: PROOFS

### A.1 Proof of Theorem 1

Let us compute the covariance matrix of the moving sums of window size $w$ (from hereon we neglect the subscript in $Y_w(t)$. The variance of a sum is easily obtained by

$$
\begin{aligned}
\sigma_{Y(t)} &= \mathrm{var}[Y(t)] = \mathrm{var}(\sum_{i=t}^{t+w-1} X_i) = \sum_{i=t}^{t+w-1} \mathrm{var}\, X_i + 2 \sum_{i<j\subset[t,t+w-1]} \mathrm{cov}(X_i, X_j) \\
&= w\sigma^2 + 2\sigma^2 \sum_{i<j\subset[t,t+w-1]} \rho_{ij} \\
&= \sigma^2 \left( w + 2 \sum_{i<j\subset[t,t+w-1]} \rho_{ij} \right).
\end{aligned} \tag{A.1}
$$

We may derive the covariance for any two sums in the process $Y(t)$ and $Y(t+g)$, $1 < g < n - w - t + 1$ as follows. Since

$$
\mathrm{var}[Y(t) + Y(t+g)] = \mathrm{var}[Y(t)] + \mathrm{var}[Y(t+g)] + 2\,\mathrm{cov}[Y(t), Y(t+g)],
$$

then

$$
\begin{aligned}
\sigma_{Y(t),Y(t+g)} &= \mathrm{cov}\left[Y(t), Y(t+g)\right] \\
&= \frac{1}{2}\left\{ \mathrm{var}\left[Y(t) + Y(t+g)\right] - \mathrm{var}\left[Y(t)\right] - \mathrm{var}\left[Y(t+g)\right] \right\}
\end{aligned} \tag{A.2}
$$

The last two variance terms within the curly brackets of (A.2) may be evaluated directly by using (A.1). Now note that the first variance term refers too to a moving sum, but the number of summands varies in accordance to the length of overlap of $Y(t)$ and $Y(t+g)$, as seen in Figure A.1. Define regions A and C as the first and second non-overlapping regions of the two windows, and B as the overlapping region. Thus $A = [t, t+g-1]$, $B = [t+g, t+w-1]$ and $C = [t+w, t+g+w-1]$, and B is of positive length only when $g < w$. Then, for any $g$,

$$
\begin{aligned}
\mathrm{var}\left[Y(t) + Y(t+g)\right] &= \mathrm{var}\left[ \sum_{i\subset A,B} X_i + \sum_{i\subset B,C} X_i \right] \\
&= \mathrm{var}\left[ \sum_{i\subset A} X_i + 2\sum_{i\subset B} X_i + \sum_{i\subset C} X_i \right]
\end{aligned} \tag{A.3}
$$

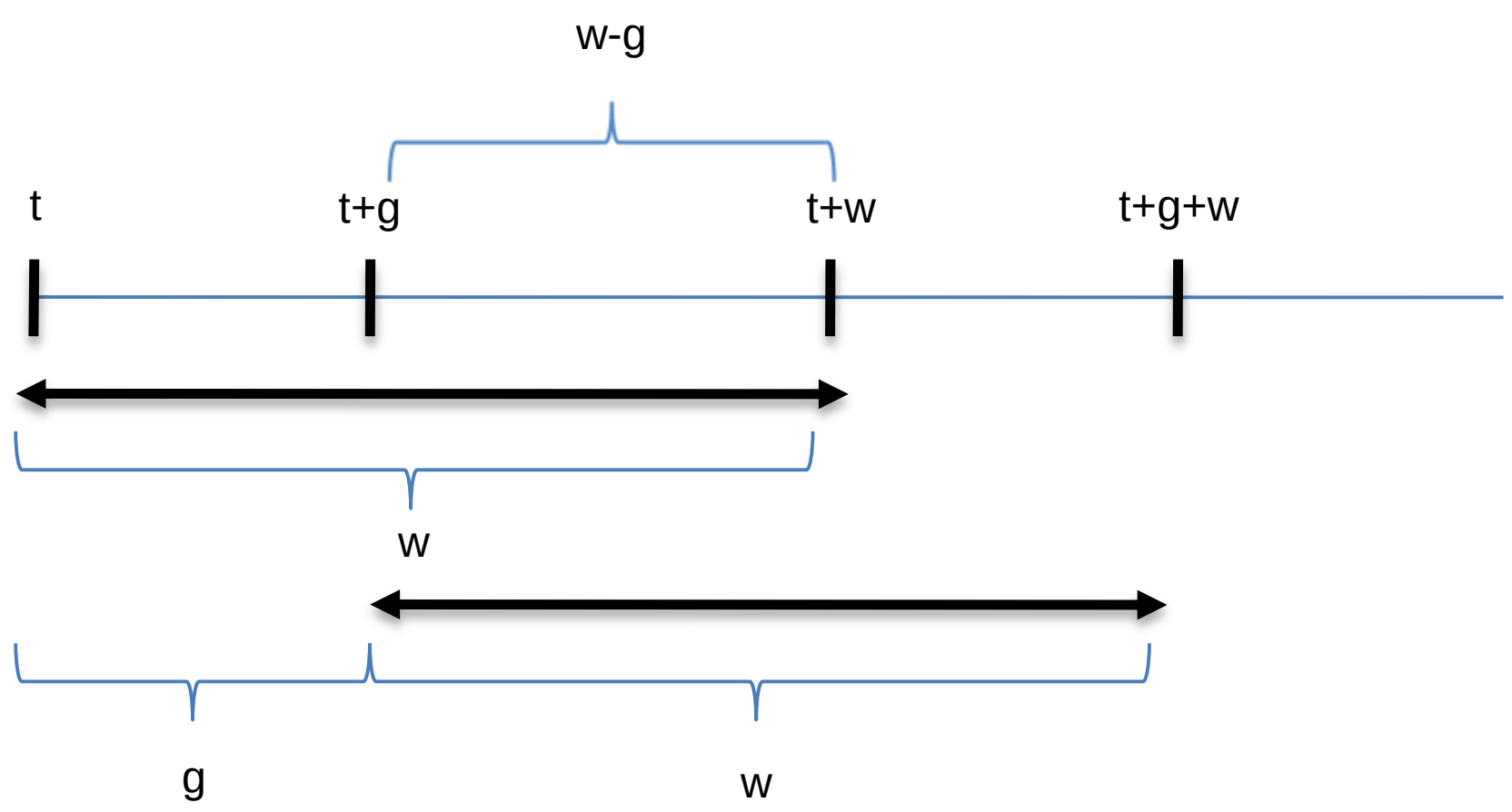

Figure A.1: Schematic description of two overlapping windows on a random process. The first window starts at point *t*, the window size is *w* and the overlapping region is of length *w-g*.

Summing directly all terms in the variance-covariance matrix of $Y(t)$ and $Y(t+g)$, (A.3) can be expressed as

$$\begin{aligned}
\operatorname{var}\left[Y(t)+Y(t+g)\right] &= 2w\sigma^2 \\
&+\sigma^2\left[2\sum_{i<j\subset A}\rho_{ij}+8\sum_{i<j\subset B}\rho_{ij}+2\sum_{i<j\subset C}\rho_{ij}\right. \\
&\qquad\left.+4\sum_{\substack{i\subset A\\ j\subset B}}\rho_{ij}+2\sum_{\substack{i\subset A\\ j\subset C}}\rho_{ij}+4\sum_{\substack{i\subset B\\ j\subset C}}\rho_{ij}+2(w-g)\right] \\
&=2\sigma^2\left[\sum_{i<j\subset A}\rho_{ij}+\sum_{i<j\subset C}\rho_{ij}+\sum_{\substack{i\subset A\\ j\subset C}}\rho_{ij}-g\right. \\
&\qquad\left.+2\left(2\sum_{i<j\subset B}\rho_{ij}+\sum_{\substack{i\subset A\\ j\subset B}}\rho_{ij}+\sum_{\substack{i\subset B\\ j\subset C}}\rho_{ij}+w\right)\right].
\end{aligned} \tag{A.4}$$

For the case of overlapping windows, such that $g<w$, by using (A.1) and (A.4), (A.2) results in

$$\operatorname{cov}\left[Y(t),Y(t+g)\right]=$$

$$\frac{1}{2}\left\{2\sigma^2\left[\sum_{i<j\subset A}\rho_{ij}+\sum_{i<j\subset C}\rho_{ij}+\sum_{\substack{i\subset A\\ j\subset C}}\rho_{ij}-g+2\left(2\sum_{i<j\subset B}\rho_{ij}+\sum_{\substack{i\subset A\\ j\subset B}}\rho_{ij}+\sum_{\substack{i\subset B\\ j\subset C}}\rho_{ij}+w\right)\right]\right.$$

$$\left.-\sigma^2\left(w+2\sum_{i<j\subset[t,t+w-1]}\rho_{ij}\right)-\sigma^2\left(w+2\sum_{i<j\subset[t+g,t+g+w-1]}\rho_{ij}\right)\right\}$$

$$=\sigma^2\left[\sum_{i<j\subset A}\rho_{ij}+\sum_{i<j\subset C}\rho_{ij}+\sum_{\substack{i\subset A\\ j\subset C}}\rho_{ij}-g+2\left(2\sum_{i<j\subset B}\rho_{ij}+\sum_{\substack{i\subset A\\ j\subset B}}\rho_{ij}+\sum_{\substack{i\subset B\\ j\subset C}}\rho_{ij}+w\right)\right.$$

$$\left.-\frac{w}{2}-\sum_{i<j\subset[t,t+w-1]}\rho_{ij}-\frac{w}{2}-\sum_{i<j\subset[t+g,t+g+w-1]}\rho_{ij}\right]$$

$$=\sigma^2\left[\sum_{i<j\subset A}\rho_{ij}+\sum_{i<j\subset C}\rho_{ij}+\sum_{\substack{i\subset A\\ j\subset C}}\rho_{ij}+2\left(2\sum_{i<j\subset B}\rho_{ij}+\sum_{\substack{i\subset A\\ j\subset B}}\rho_{ij}+\sum_{\substack{i\subset B\\ j\subset C}}\rho_{ij}\right)\right.$$

$$\left.-\sum_{i<j\subset[t,t+w-1]}\rho_{ij}-\sum_{i<j\subset[t+g,t+g+w-1]}\rho_{ij}+w-g\right] \tag{A.5}$$

In the case of non-overlapping windows, such that $g\geq w$, by using (A.1), (A.3) reduces to

$$\operatorname{var}\left[Y(t)+Y(t+g)\right]=\operatorname{var}\left(\sum_{i=t}^{t+w-1}X_i+\sum_{i=t+g}^{t+g+w-1}X_i\right)$$

$$=\sigma^2\left[2w+2\left(\sum_{i<j\subset[t,t+w-1]}\rho_{ij}+\sum_{i<j\subset[t+g,t+g+w-1]}\rho_{ij}+\sum_{\substack{i\subset[t,t+w-1]\\ j\subset[t+g,t+g+w-1]}}\rho_{ij}\right)\right]$$

$$=2\sigma^2\left(w+\sum_{i<j\subset[t,t+w-1]}\rho_{ij}+\sum_{i<j\subset[t+g,t+g+w-1]}\rho_{ij}+\sum_{\substack{i\subset[t,t+w-1]\\ j\subset[t+g,t+g+w-1]}}\rho_{ij}\right)$$

and thus (A.5) becomes

$$\begin{aligned}
&\operatorname{cov}\left[Y(t),Y(t+g)\right] \\
&=\frac{1}{2}\left[2\sigma^2\left(w+\sum_{i<j\subset[t,t+w-1]}\rho_{ij}+\sum_{i<j\subset[t+g,t+g+w-1]}\rho_{ij}+\sum_{\substack{i\subset[t,t+w-1]\\ j\subset[t+g,t+g+w-1]}}\rho_{ij}\right)\right. \\
&\qquad\left.-\sigma^2\left(w+2\sum_{i<j\subset[t,t+w-1]}\rho_{ij}\right)-\sigma^2\left(w+2\sum_{i<j\subset[t+g,t+g+w-1]}\rho_{ij}\right)\right] \\
&=\sigma^2\sum_{\substack{i\subset[t,t+w-1]\\ j\subset[t+g,t+g+w-1]}}\rho_{ij}.
\end{aligned} \tag{A.6}$$

## A.2 Proof of Corollary 1

Consider auto-correlation between $X_1,...,X_n$, such that $\rho_{ij}=\rho^{|j-i|}$ for any $i\neq j$. Note that in this case, for interval $L$ of length $l$,

$$\sum_{i<j\subset L}\rho_{ij}=\sum_{i<j\subset L}\rho^{|j-i|}=(l-1)\rho+(l-2)\rho^2+...+(l-(l-1))\rho^{l-1}=\sum_{i=1}^{l-1}(l-i)\rho^i,$$

and for two non-overlapping windows $L$ and $L^{'}$ of length $l$ and $l^{'}$, respectively, with distance $d$ between their starting points, $d\geq l$, $\sum_{\substack{i\subset L\\ j\subset L^{'}}}\rho_{ij}=\sum_{\substack{i\subset L\\ j\subset L^{'}}}\rho^{|j-i|}=\sum_{i=1}^{l}\sum_{j=1}^{l^{'}}\rho^{|d+j-i|}$ .

Thus (A.1) becomes

$$\operatorname{var}\left[Y(t)\right]=\sigma^2\left(w+2\sum_{i=1}^{w-1}(w-i)\rho^i\right), \tag{A.7}$$

(A.5) becomes

$$\begin{aligned}
&\operatorname{cov}\left[Y(t),Y(t+g)\right]= \\
&=\sigma^2\left[\sum_{i=1}^{g-1}(g-i)\rho^i+\sum_{i=1}^{g-1}(g-i)\rho^i+\sum_{i=1}^{g}\sum_{j=1}^{g}\rho^{|w+j-i|}\right. \\
&\qquad+2\left(2\sum_{i=1}^{w-g-1}(w-g-i)\rho^i+\sum_{i=1}^{g}\sum_{j=1}^{w-g}\rho^{|g+j-i|}+\sum_{i=1}^{w-g}\sum_{j=1}^{g}\rho^{|w-g+j-i|}\right) \\
&\qquad\left.-\sum_{i=1}^{w-1}(w-i)\rho^i-\sum_{i=1}^{w-1}(w-i)\rho^i+w-g\right] \\
&=\sigma^2\left[2\sum_{i=1}^{g-1}(g-i)\rho^i-2\sum_{i=1}^{w-1}(w-i)\rho^i+\sum_{i=1}^{g}\sum_{j=1}^{g}\rho^{|w+j-i|}\right. \\
&\qquad\left.+2\left(2\sum_{i=1}^{w-g-1}(w-g-i)\rho^i+\sum_{i=1}^{g}\sum_{j=1}^{w-g}\rho^{|g+j-i|}+\sum_{i=1}^{w-g}\sum_{j=1}^{g}\rho^{|w-g+j-i|}\right)+w-g\right] \\
&=\sigma^2\left[\sum_{i=1}^{g}\sum_{j=1}^{g}\rho^{|w+j-i|}+2\left(2\sum_{i=1}^{w-g-1}(w-g-i)\rho^i+\sum_{i=1}^{g}\sum_{j=1}^{w-g}\rho^{|g+j-i|}\right.\right. \\
&\qquad\left.\left.+\sum_{i=1}^{w-g}\sum_{j=1}^{g}\rho^{|w-g+j-i|}+\sum_{i=1}^{g-1}(g-i)\rho^i-\sum_{i=1}^{w-1}(w-i)\rho^i\right)+w-g\right]
\end{aligned} \tag{A.8}$$

and (A.6) becomes

$$\operatorname{cov}\left[Y(t),Y(t+g)\right]=\sigma^2\sum_{i=1}^{w}\sum_{j=1}^{w}\rho^{|g+j-i|}. \tag{A9}$$

Specifically in the case of adjacent windows, such that $g=w$, (A.12) can be expressed with no regard to *g*, and $\rho$ takes the power $|w+j-i|$.

## A.3 Proof of Corollary 2

Consider $\Sigma_X$ such that common correlation exists between $X_1,...,X_n$, $\rho_{ij}=\rho$, $-1\le\rho\le 1$, for any $i\ne j$. Note that in this case, for interval $L$ of length $l$, $\sum_{i<j\subset L}\rho_{ij}=\frac{l(l-1)}{2}\rho$, and for two non-overlapping windows $L$ and $L'$ of length $l$ and $l'$, respectively, $\sum_{\substack{i\subset L\\ j\subset L'}}\rho_{ij}=l\cdot l'\rho$. Then (A.1) becomes

$$\operatorname{var}[Y(t)]=\sigma^2\left(w+2\frac{w(w-1)}{2}\rho\right)=\sigma^2 w[1+(w-1)\rho], \tag{A.10}$$

and thus (A.5) becomes

$$\begin{aligned}
&\operatorname{cov}\left[Y(t),Y(t+g)\right]=\\
&=\sigma^2\left[2\frac{g(g-1)}{2}\rho+g^2\rho+2\left(2\frac{(w-g)(w-g-1)}{2}\rho+2g(w-g)\rho\right)\right.\\
&\qquad\left.-2\frac{w(w-1)}{2}\rho+w-g\right]\\
&=\sigma^2\left[\rho\Big(g(g-1)+g^2+2(w-g)(w-g-1)+4g(w-g)-w(w-1)\Big)+w-g\right]\\
&=\sigma^2\left[\rho\left\{g(2g-1)+2(w-g)(w+g-1)-w(w-1)\right\}+w-g\right]
\end{aligned} \tag{A.11}$$

and (A.6) becomes

$$\operatorname{cov}[Y(t),Y(t+g)]=\sigma^2\sum_{\substack{i\subset[t,t+w-1]\\ j\subset[t+g,t+g+w-1]}}\rho_{ij}=\sigma^2 w^2\rho. \tag{A.12}$$

Specifically in the case of adjacent windows, such that $g=w$, the *j* index takes the values $[t+w,\ \ t+2w-1]$.

## APPENDIX B: VERIFICATION OF FORMULAS

Covariance and tail probabilities based on the formulas developed in Section 2 can be confirmed by comparing them to estimates based on simulated random processes. In the $i^{th}$ realization, $i = 1,...,N$, sequences of observations denoted $X^{*i}_{j1},...,X^{*i}_{jn}$ are repeatedly sampled $J$ times, $j = 1,...,J$, from a given multi-normal distribution $F^0_X = MVN(\bar{\mu}_0, \Sigma_X)$, with all entries in $\bar{\mu}_0$ equal $\mu_0$ and all entries in the diagonal of $\Sigma_X$ equal $\sigma_X$. Then, the corresponding $J$ processes of moving sums $Y^{*i}_{jw}(1),...,Y^{*i}_{jw}(n+w-1)$ are calculated for a given window size *w*, and denote their covariance matrix by $\Sigma^{*i}_Y$. A scan statistic $S^{*i}_{jw}$ is calculated for each of the *J* processes of moving sums, along with the proportion of observations larger than a given value *s*, $p^{*i}_w(s) = \sum_{j=i}^{J} \frac{I(S^{*i}_{jw} > s)}{J}$. The procedure is repeated *N* times and the observed covariance matrices are averaged to produce an estimate of $\Sigma_Y$,

$$\hat{\sigma}_{Y(t)} = \bar{\sigma}^*_{Y(t)} = \frac{1}{N}\sum_{i=1}^{N}\sigma^{*i}_{Y(t)}, \tag{B.1}$$

$$\hat{\sigma}_{Y(t),Y(t+g)} = \bar{\sigma}^*_{Y(t),Y(t+g)} = \frac{1}{N}\sum_{i=1}^{N}\sigma^{*i}_{Y(t),Y(t+g)} \tag{B.2}$$

and an estimate of $\Pr_{F^0_S}(S_w > s)$,

$$\hat{p}_w(s) = \bar{p}^*_w(s) = \frac{1}{I}\sum_{i=1}^{N}\sum_{j=i}^{J}\frac{I(S^{*i}_{jw} > s)}{J}, \tag{B.3}$$

which is in effect the overall proportion of values larger than *s*.

Here *J* and *N* were both set to 1000, enabling the outcome of stable estimates. The process length parameter, *n*, was set to 7, and *w*, the window size, was set to 3. These values allowed both overlapping and non-overlapping windows, with relatively short computer runtime. Equivalent results were obtained for other values of *n* and *w*.

For simplicity, $\mu_0$ was set to 0 and $\sigma_X$ was set to 1. For a common covariance structure, $\rho$ was assigned values greater than $\frac{-1}{n-1}$, here -0.14, in order to guarantee positive definiteness of the covariance matrix. For an auto-covariance structure, $\rho$ was given values within the range [-1,1]. Note that the cases of $\rho = 0$ and $\rho = 1$ yield the same matrices for auto-covariance and common structures, but results are presented only within the common covariance part. For a general covariance structure of the original normal process, correlation values were independently sampled from $U[-1,1]$. Whenever the resulting sampled matrix was not positive definite, Higham's algorithm was employed to compute the nearest positive definite matrix (Higham 2002 and Cheng and Higham 1998) using the Matrix R package (Bates and Maechler 2010). The equivalence of the exact and estimated matrices is detailed in Table B.1 for all specified correlation structures.

The results presented for the general structure case are based on the covariance matrix:

$$\Sigma_X = \begin{pmatrix} 1.000 & -0.314 & -0.454 & -0.154 & -0.107 & 0.395 & 0.650 \\ -0.314 & 1.000 & 0.050 & 0.452 & 0.095 & 0.474 & -0.230 \\ -0.454 & 0.050 & 1.000 & 0.110 & 0.538 & -0.342 & 0.210 \\ -0.154 & 0.452 & 0.110 & 1.000 & -0.359 & -0.045 & -0.127 \\ -0.107 & 0.095 & 0.538 & -0.359 & 1.000 & -0.312 & -0.035 \\ 0.395 & 0.474 & -0.342 & -0.045 & -0.312 & 1.000 & 0.495 \\ 0.650 & -0.230 & 0.210 & -0.127 & -0.035 & 0.495 & 1.000 \end{pmatrix} \quad \text{(B.4)}$$

and the obtained exact covariance matrix for the corresponding moving sums is, as calculated by (A.1), (A.5) and (A.6), is

$$\Sigma_Y = \begin{pmatrix} 1.564801 & 1.740606 & 1.529404 & 1.459909 & 1.681448 \\ 1.740606 & 4.22419 & 2.995086 & 1.92215 & 0.213491 \\ 1.529404 & 2.995086 & 3.576884 & 1.23134 & 0.528563 \\ 1.459909 & 1.92215 & 1.23134 & 1.569523 & 1.306504 \\ 1.681448 & 0.213491 & 0.528563 & 1.306504 & 3.297022 \end{pmatrix}$$

while the obtained simulation estimate for this matrix is

$$\Sigma_Y^* = \begin{pmatrix} 1.564903 & 1.738066 & 1.52519 & 1.460384 & 1.684061 \\ 1.738066 & 4.216156 & 2.986097 & 1.920295 & 0.214572 \\ 1.52519 & 2.986097 & 3.568333 & 1.22792 & 0.526874 \\ 1.460384 & 1.920295 & 1.22792 & 1.570396 & 1.308937 \\ 1.684061 & 0.214572 & 0.526874 & 1.308937 & 3.302989 \end{pmatrix}$$

with standard errors all smaller than 0.006.

Table B.1: Exact and estimated tail probability and covariance, by covariance structure. Covariance is given for a process of length 7, window size 3 and $t$=1, for overlapping windows ($g$=2) and non-overlapping windows ($g$=4). Simulation was repeated with $N$=1000 and $J$=1000. In parenthesis - standard error for the estimate. Equality of exact and estimated results was consistent for other values of process parameters. The tail probability is calculated for an "observed" scan statistic $s$=3

| Covariance structure | $\rho$ | $p_w(s)$ | $\hat{p}_w(s)$ | $\sigma_{Y(t),Y(t+g)}$ | $\hat{\sigma}_{Y(t),Y(t+g)}$ | $\sigma_{Y(t),Y(t+g)}$ | $\hat{\sigma}_{Y(t),Y(t+g)}$ |
|---|---|---|---|---|---|---|---|
| | | | | $g$=2 (overlapping) | | $g$=4 (non-overlapping) | |
| General | See (B.4) | 0.14059 | 0.14031 (0.00036) | 1.5294 | 1.5252 (0.0028) | 1.6814 | 1.6841 (0.0028) |
| Common | -0.1 | 0.10840 | 0.10813 (0.00031) | 0.2 | 0.1981 (0.0024) | -0.9 | -0.8999 (0.0026) |
| | 0 | 0.14541 | 0.14512 (0.00034) | 1 | 0.9938 (0.0031) | 0 | -0.0034 (0.0029) |
| | 0.1 | 0.16978 | 0.17001 (0.00038) | 1.8 | 1.8039 (0.0040) | 0.9 | 0.9027 (0.0038) |
| | 0.25 | 0.19132 | 0.19174 (0.00038) | 3 | 2.9945 (0.0053) | 2.25 | 2.2451 (0.0050) |
| | 0.5 | 0.20520 | 0.20611 (0.00040) | 5 | 5.0079 (0.0075) | 4.5 | 4.5063 (0.0073) |
| | 0.75 | 0.20277 | 0.20278 (0.00039) | 7 | 7.0095 (0.0103) | 6.75 | 6.7579 (0.0103) |
| | 1 | 0.15866 | 0.15830 (0.00037) | 9 | 0.89920 (0.0127) | 9 | 0.89920 (0.0127) |
| Auto | -1 | 0.00270 | 0.00272 (0.00005) | 1 | 0.9978 (0.0015) | 1 | 0.9978 (0.0015) |
| | -0.75 | 0.01078 | 0.01086 (0.00009) | 0.6606 | 0.6605 (0.0013) | 0.3713 | 0.3724 (0.0012) |
| | -0.5 | 0.03317 | 0.03315 (0.00017) | 0.5625 | 0.5623 (0.0015) | 0.1406 | 0.1413 (0.0015) |
| | -0.25 | 0.08112 | 0.08169 (0.00027) | 0.6602 | 0.6621 (0.0022) | 0.0413 | 0.0445 (0.0021) |
| | -0.1 | 0.11900 | 0.11901 (0.00032) | 0.8281 | 0.8292 (0.0028) | 0.0083 | 0.0090 (0.0027) |
| | 0.1 | 0.17095 | 0.17086 (0.00037) | 1.2321 | 1.2292 (0.0037) | 0.0123 | 0.0097 (0.0034) |
| | 0.25 | 0.20643 | 0.20607 (0.00040) | 1.7227 | 1.7268 (0.0045) | 0.1077 | 0.1132 (0.0043) |
| | 0.5 | 0.24851 | 0.24824 (0.00043) | 3.0625 | 3.0618 (0.0061) | 0.7656 | 0.7635 (0.0055) |
| | 0.75 | 0.25700 | 0.25698 (0.00044) | 5.3477 | 5.3478 (0.0090) | 3.0080 | 3.0154 (0.0078) |